============cut here==============
\documentstyle[preprint,prb,aps]{revtex}
\begin{document}
\draft
\title{
The Shape of Bucky Onions
}
\author{Jian Ping Lu$^{\mbox{\dag}}$ and Weitao Yang$^{\mbox{\ddag}}$}
\address{
$^{\mbox{\dag}}$Department of Physics and Astronomy,
University of North Carolina at Chapel Hill,
Chapel Hill, North Carolina 27599 \\
$^{\mbox{\ddag}}$ Department of Chemistry, Box 90354,
Duke University, Durham, NC 27708-0354
}
\date{\today}
\maketitle

\begin{abstract}
{\bf The morphology of Bucky onions is investigated by
{\it ab initio} calculations using Yang's new $O(N)$ method.
It is found that for large single shell
fullerenes with $I_h$ symmetry, the spherical morphology
has the lower energy than that of polyhedrons.
The formation energy per atom follows a simple scaling
law. Including an estimate of inter-shell van de Waals
interactions leads to the conclusion
that spherical multiple shell Bucky onions
are likely the most
stable structure of large carbon clusters.
These results are in good agreement with
recent observation.
}
\end{abstract}


The exciting discovery of concentric spherical graphitic shells
by Ugarte\cite{ugarte} raises several intriguing questions.
Most fundamentally, it suggests that multi-shell
fullerenes --- so called Bucky onions --- are the most stable structures
of finite carbon clusters.
This, and the remarkable spherical shape of these onions call for theoretical
studies. For cluster size up to $N=240$ carbon atoms, Adams {\it et
al.}\cite{adams}
have performed first principle calculations and showed
that ``ball shaped'' fullerenes have a lower energy
than that of tube-shaped. However,
based either on empirical potentials calculations, geometric
consideration and elastic theory,
it has been concluded by several groups that for
large fullerenes the morphology is
polyhedrally faceted\cite{mckay,yoshida,maiti,tersoff}.
This appears to contradict the experimental observation
of spherical shells.

In this letter, we report results of first
principle calculations for carbon clusters up to $1,000$
atoms by the ``divide-and-conquer'' method of Yang \cite{yang}.
These are the first {\it ab initio}
calculations for such large molecules.
The results unequivocally demonstrate that for single
shell fullerenes the spherical morphology has a lower energy
than those of polyhedrons. Furthermore we show that the leading
$N$ dependence of the total energy obtained can be
understood in term of the H\"{u}ckel theory. Including the
inter-shell van de Waals interactions leads to the conclusion
that the multi-shell concentric spherical Bucky onions
are likley the most stable structure of large carbon clusters,
as observed in recent experiments\cite{ugarte}.

Most methods of first principle calculations are limited
to molecules of small or modest size. Recently one of us has developed
an efficient $O(N)$ algorithm\cite{yang},
designed specifically for large
molecules. The method has been implemented for
general molecular computations and tested
in small systems against the traditional Kohn-Sham
density-functional
theory. The tests performed were molecular bonding energetic, molecular
internal rotation potential barriers, and
total electronic density of states \cite{yang}.
In our present calculations, a cluster is divided into
subsystems with one atom each, and the local basis set for each subsystem
used includes the atomic orbitals of the atom as well as those up
to the third nearest neighbors. We use the local density
approximation(LDA)
for the exchange and correlation energy
and numerical LDA solutions of a
spherical carbon atom for the atomic orbitals.
To reduce the computational effort the non-selfconsistent Harris
functional \cite{harris} is employed as well as the $I_h$ symmetry.
Using this new method we are able to carry out
{\it ab initio} calculations for carbon clusters up
to $1,000$ atoms. Selective results are summarized
in Table I.

We restrict our investigation to Goldberg type I
fullerenes\cite{yoshida},
in which the number of atoms in $n$th shell is $N(n)=60n^2$ and
the average radius can be approximated
by $\bar{R}_n\approx 2.4\bar{b}n$, where $\bar{b}$ is the average C-C
bond length. Thus, the inter-shell spacing is close to the
inter-layer distance in the graphite, giving rise to the  maximum
inter-shell van de Waals attraction.
For the smallest member $C_{60}$, all sites are equivalent,
the shape can be classified equally well as spherical or polyhedral.
The next shell $C_{240}$ has three independent atomic
sites (Fig.1). The $I_h$ symmetry reduces
the independent variables to seven.
A global search for the minimum is out of reach
of our present computational capabilities.
We have investigated four most likely morphologies. 1)
Sphere (S), in which all three sites
are assumed to be on a spherical surface.
2) Icosahedrally (I) faceted, here all three sites
are assumes to be on a faceted surface. A generalization of this
morphology to large fullerenes is an Icosahedron\cite{maiti}.
3) Truncated Icosahedron (TI), this is an inflated version of $C_{60}$,
in which site 1 and 2 (Fig.1) are assumed to be on the pentagonal facet
while 2 and 3 are on a hexagonal facet.
4) The low energy morphology obtained by Yoshida and Osawa (YS)\cite{yoshida}.
This morphology is neither spherical nor simply faceted. However, as measured
by the deviation from a sphere it is very close to (TI).

For the morphologies (S), (I) and (TI) the total energy was minimized
with respect to two independent bond lengths (Fig.1). For
the (YS) shape we take the ratios of various bond
lengths to be the same as that given by Ref.4.
Table I lists the formation energy and corresponding geometric
parameters obtained. As one can see,
the formation energy for S-$C_{240}$
is significantly smaller than that of I-$C_{240}$,
TI-$C_{240}$ and YO-$C_{240}$.
Therefore, our results demonstrate that for $C_{240}$
the spherical morphology has a lower energy than that of polyhedrons.
This conclusion remains unchanged when the calculations were
repeated using the atomic orbitals from up to fourth
nearest neighbor atoms.

Included in Table I are also results of similar
investigations for $C_{540}$. Again it is found that
the spherical morphology has significant lower energy.
Strong evidence in favor of the spherical shape is the
fact that in all cases the formation energy is correlated
with the sphericity as measured
by the standard deviation (SD) from a perfect sphere (Table I).
The more spherical is the morphology the lower is the energy.
Based on these results we conclude
that {\it for large single-shell fullerenes the spherical shape is the
preferred morphology}.

To understand the $N$ dependence of the formation energy
several more clusters are examined.
Plotted in Fig.2 are the formation energy per atom
$E_c$ versus $1/N$. $E_c$ decreases monotonically
with increasing fullerene size $N$.
A straight line fitting to the four largest fullerenes
calculated leads to
\begin{equation}
E_c(N)=-7.178(1.0-\frac{4.69}{N}) \;\; eV \; .
\end{equation}
Thus, $E_c(\infty)=-7.178$eV is the extrapolated cohesive energy
per atom for single graphitic layer.
This number is close to that found by other
calculations and by experiments\cite{cohen}.

The simple $1/N$ dependence can be understood in terms of
the H\"{u}ckel theory. In the nearest neighbor (n.n.)
tight binding calculation there are two contributions
to the energy difference between a fullerene and
that of a graphitic layer.
The first is due to the finite size of the fullerene.
Exact calculation shows that the tight binding energy
increases linearly with $1/N$,
$E_t(N)=-1.574t(1-\frac{0.835}{N})$, where $t$
is the n.n. hoping integral \cite{note1}.
This relation should reflects the
approximate $N$ dependence of $\sigma$ bonding energy.
The second, and more important contribution comes
from the fact that for $\pi$ orbitals the n.n. overlaps
depends on the molecular size\cite{adams}. For spherical morphology
$t_\pi$ is proportional to the $cos(\phi)$, where $\phi$
is the radial angle spanned by the nearest neighbor C-C bond,
$t_\pi=t_o\cos{\phi}=t_o\cos(2\arcsin(\frac{b}{2R_N}))$
where $R_N$ is the radius of the sphere and $t_o$
is the limiting value for the graphitic plane.
Taking $R_N\approx 2.4b\sqrt{N/60}$ and expanding in $1/N$ to
the leading order one gets
\begin{equation}
t_\pi \approx t_o(1.0-\frac{5.21}{N})
\end{equation}
Compared with Eq.1 one sees the dominant contribution to
the $N$ dependence of the cohesive energy is due to the change
in $p_\pi$ orbital overlaps.

We now discuss the energetics of multi-shell Bucky
onions and their morphology. Here
one needs to include the inter-shell interactions.
In Goldberg type I fullerenes
the spacing between successive shells, $\sim 3.4$\AA, is almost the
same as that of interlayer spacing in graphite.
Therefore one expects that interactions are dominated by the
van de Waals force. In solid $C_{60}$, the intermolecular
interactions have been successfully modeled using the
van de Waals interactions\cite{lu2}.
There it was found that 90\% of cohesion energy can be calculated
by simply summing over all the inter-fullerenes C-C interactions using the
Lennard-Jones potential. For Bucky onions, this method should be
an even better approximation as the inter-shell
spacing are uniform. Therefore we use the
same Lennard-Jones potential
($\epsilon=2.964$ meV, $\sigma=3.407\AA$)
to evaluate the inter-shell interactions.

The fact that $R_{n+1}-R_n \sim \sigma$ suggests that the inter-shell
attraction is the largest when both shells have the spherical shape.
Indeed numerical calculations show that the attractive
interactions in the sphere morphology is significantly
larger than that of polyhedrons,
because the non-uniform spacing between shells in the latter.
For example, the attraction between
$C_{240}$ and S-$C_{540}$ (-17.7 eV, see Table II) is
significantly larger than that between I-$C_{240}$ and I-$C_{540}$ (-12.3 eV).
Thus, the inter-shell
interactions also favor the spherical shape.
Therefore we conclude that
{\em the shape of a multi-shell Bucky onion is spherical}.
This is in agreement with experimental observation of spherical
shaped multi-shell concentric Bucky onions\cite{ugarte}.
In Fig.3 a non-perspective view of a
five shell Bucky onion is shown.

{}From Table II the van de Waals interactions between the $n$th
and $(n+1)$th shell is found to be well described by
$V_{n,n+1}=-3.0n(n+1)$ eV\cite{note2}.
Using the extrapolated energy Eq.1 for
the single shell fullerene, one can estimates that when $n\geq3$
two shell Bucky onion with $N(n)+N(n+1)$ atoms is more stable
than a single shell fullerene with the same total number
of atoms. The first case where the number $N(n)+N(n+1)$ also
belongs to a type I Goldberg fullerene is when $n=3$. Thus our
calculations suggest that energetically $C_{(540,960)}$ is more
stable than $C_{1500}$\cite{note3}. This result is rather
different from earlier calculation where the crossover was
found to occur at $N \sim 6000$ \cite{maiti}.
Finally, it is clear that for large clusters the
dense Bucky onions are most stable. This is because
the van de Waals attraction is $\sim l^3$ for a $l$ shell onion,
while the cost is $\sim l$ compared with a single shell fullerene.

In conclusion, we have shown that the shape of large
fullerenes are spherical regardless whether it is single
or multiple shell. By including the van de Waals
interactions between shells it is demonstrated
that spherical dense Bucky onions are likely the most stable
structure for large carbon clusters. This is in good
agreement with recent experimental observations of these
concentric spherical shells. The possibility of
forming a lattice of Bucky onions should open
up a new area of fullerene research and
many interesting physical properties can be expected.

\acknowledgments
We thank A. Maiti and J. Bernholc for communicating their
results prior to the publication.
This work is supported by The University Research Council of
the University of North Carolina at Chapel Hill (JPL),
by the National Science Foundation and
North Carolina Supercomputing Center (WY).

\begin{figure}

\caption{Asymmetric part of large fullerene structures.
(a) $C_{240}$, shown are three independent sites and
two bond labeling (see text). (b) Similar drawing for $C_{540}$.
}

\bigskip
\bigskip

\caption{The formation energy vs $1/N$. The straight line
(Eq.1) is the least square fit into the four points
$N=180, 240, 540, 960$.}

\bigskip
\bigskip

\caption{A non-perspective view of the five shells dense Bucky onion
$C_{(60,240,540,960,1500)}$. For clarity, only portion of each shell is shown.
}

\end{figure}

\begin{table}

\caption{Formation energies per atom for various fullerenes with different
morphologies:
S -- spherical, I -- Icosahedron faceted, TI -- truncated Icosahedron,
YO -- the structure given in Ref.4.
$\bar{b}$ -- average bond length,
$\bar{R}$ -- Average radius, SD -- standard deviation from a perfect sphere. }

\begin{tabular}{lllll}
 & $E_c$ (eV)	& $\bar{b}$(\AA) ($b_1$, $b_2$) & $\bar{R}$(\AA) & SD (\AA) \\
\tableline
\tableline
S-$C_{240}$	&   -7.05	&      1.43 (1.43, 1.43)	& 7.12	& 0.00 \\
I-$C_{240}$	&   -6.90   	&      1.45 (1.46, 1.42)	& 7.27	& 0.39 \\
TI-$C_{240}$	&   -6.92   	&      1.46 (1.47, 1.43)	& 7.09	& 0.18 \\
YO-$C_{240}$ 	&   -6.97   	&      1.45 (1.45, 1.40)    	& 7.03	& 0.17 \\
\tableline
\tableline
S-$C_{540}$  	&   -7.09   	&      1.41 (1.42, 1.41)	& 10.5	& 0.00	\\
I-$C_{540}$ 	&   -6.86   	&      1.43 (1.44, 1.40)	& 10.4	& 0.52 \\
\end{tabular}

\end{table}

\begin{table}

\caption{The van de Waals interactions between the first five shells
of a dense spherical Bucky onion.}

\begin{tabular}{lllll}
$V_{ij}$ (eV)	&   $C_{240}$      & $C_{540}$    & $C_{960}$    & $C_{1500}$ \\
\tableline
\tableline
$C_{60}$	&   -5.57 &  -0.81 &  -0.20 &  -0.07 \\
$C_{240}$       &      	  &  -17.7 &  -2.34 &  -0.55 \\
$C_{540}$  	&	  &	   &  -35.9 &  -4.28 \\
$C_{960}$	&	  &	   &	    &  -59.9 \\
\end{tabular}

\end{table}


\begin{references}
\bibitem{note1}
If all nearest neighbor hoping are identical, caculations show
that all type I
Goldberg series are close-shelled, namely
the HOMO are completely occupied. This is generally not true
for type II in which $N(n)=20n^2$.
\bibitem{note2}
In the case of spherical morphology, it is found that the
rotational barrier of inner shell vs outer shell is very small.
Thus, free rotation is expected at room temperature.
In contrast, for polyhedron morphologies the barriers are very high.
\bibitem{note3}
The next occurrence for the identity $n^2+(n+1)^2=m^2$
is when $n=20,m=29$. In this case $N=50460$,
the van de Waals attraction $V_{20,21}=1261$ eV is large and
the two-shell onion is more stable.
\end{references}
\end{document}